\documentclass[aps,prb,reprint,noshowkeys,superscriptaddress]{revtex4-1}
\usepackage{graphicx,dcolumn,bm,xcolor,microtype,multirow,amscd,amsmath,amssymb,amsfonts,physics,wrapfig,txfonts}
\usepackage[version=4]{mhchem}

\newcommand{\ie}{\textit{i.e.}}

\usepackage[normalem]{ulem}

\newcommand{\mc}{\multicolumn}

\newcommand{\tabc}[1]{\multicolumn{1}{c}{#1}}
\newcommand{\QP}{\textsc{quantum package}}

\usepackage[
	colorlinks=true,
    citecolor=blue,
    breaklinks=true
	]{hyperref}
\begin{document}

\newcommand{\LCPQ}{Laboratoire de Chimie et Physique Quantiques (UMR 5626), Universit\'e de Toulouse, CNRS, UPS, France}

\title{The performance of CIPSI on the ground state electronic energy of benzene}

\author{Pierre-Fran\c{c}ois Loos}
\email{loos@irsamc.ups-tlse.fr}
\affiliation{\LCPQ}
\author{Yann Damour}
\affiliation{\LCPQ}
\author{Anthony Scemama}
\email{scemama@irsamc.ups-tlse.fr}
\affiliation{\LCPQ}

\begin{abstract}
Following the recent work of Eriksen \textit{et al.}~[\href{https://dx.doi.org/10.1021/acs.jpclett.0c02621}{J.~Phys.~Chem.~Lett.~\textbf{11}, 8922 (2020)}], we report the performance of the \textit{Configuration Interaction using a Perturbative Selection made Iteratively} (CIPSI) method on the non-relativistic frozen-core correlation energy of the benzene molecule in the cc-pVDZ basis. Following our usual protocol, we obtain a correlation energy of $-863.4$ m$E_h$ which agrees with the theoretical estimate of $-863$ m$E_h$ proposed by Eriksen \textit{et al.}~using an extensive array of highly-accurate new electronic structure methods.
\end{abstract}

\maketitle

Although sometimes decried, one cannot deny the usefulness of benchmark sets and their corresponding reference data for the electronic structure community.
These are indeed essential for the validation of existing theoretical models and to bring to light and subsequently understand their strengths and, more importantly, their weaknesses.
In that regard, the previous benchmark datasets provided by the \textit{Simons Collaboration on the Many-Electron Problem} have been extremely valuable. \cite{Leblanc_2015,Motta_2017,Williams_2020}
The same comment applies to the excited-state benchmark set of Thiel and coworkers. \cite{Sauer_2009,Schreiber_2008,Silva-Junior_2010a,Silva-Junior_2010b,Silva-Junior_2010c}
Following a similar goal, we have recently proposed a large set of highly-accurate vertical transition energies for various types of excited states thanks to the renaissance of selected configuration interaction (SCI) methods \cite{Bender_1969,Huron_1973,Buenker_1974} which can now routinely produce near full configuration interaction (FCI) quality excitation energies for small- and medium-sized organic molecules. \cite{Loos_2018a,Loos_2019,Loos_2020a,Loos_2020b,Loos_2020c}

In a recent article, \cite{Eriksen_2020} Eriksen \textit{et al.}~have proposed a blind test for a particular electronic structure problem inviting several groups around the world to contribute to this endeavour.
In addition to coupled cluster theory with singles, doubles, triples, and quadruples (CCSDTQ), \cite{Oliphant_1991,Kucharski_1992} a large panel of highly-accurate, emerging electronic structure methods were considered:
	(i) the many-body expansion FCI (MBE-FCI), \cite{Eriksen_2017,Eriksen_2018,Eriksen_2019a,Eriksen_2019b}
	(ii) three SCI methods including a second-order perturbative correction [adaptive sampling CI (ASCI), \cite{Tubman_2016,Tubman_2018,Tubman_2020} iterative CI (iCI), \cite{Liu_2014,Liu_2016,Lei_2017,Zhang_2020} and semistochastic heat-bath CI (SHCI) \cite{Holmes_2016,Holmes_2017,Sharma_2017}],
	(iii) the full coupled-cluster reduction (FCCR) \cite{Xu_2018,Xu_2020} which also includes a second-order perturbative correction, 
	(iv) the density-matrix renornalization group (DMRG) approach, \cite{White_1992,White_1993,Chan_2011} and
	(v) two flavors of FCI quantum Monte Carlo (FCIQMC), \cite{Booth_2009,Cleland_2010} namely AS-FCIQMC \cite{Ghanem_2019} and CAD-FCIQMC. \cite{Deustua_2017,Deustua_2018}
We refer the interested reader to Ref.~\onlinecite{Eriksen_2020} and its supporting information for additional details on each method and the complete list of references.
Soon after, Lee \textit{et al.}~reported phaseless auxiliary-field quantum Monte Carlo \cite{Motta_2018} (ph-AFQMC) correlation energies for the very same problem. \cite{Lee_2020}

The target application is the non-relativistic frozen-core correlation energy of the ground state of the benzene molecule in the cc-pVDZ basis.
The geometry of benzene has been optimized at the MP2/6-31G* level \cite{Schreiber_2008} and its coordinates can be found in the supporting information of Ref.~\onlinecite{Eriksen_2020} alongside its nuclear repulsion and Hartree-Fock energies.
This corresponds to an active space of 30 electrons and 108 orbitals, \ie, the Hilbert space is of the order of $10^{35}$ Slater determinants.
Needless to say that this size of Hilbert space cannot be tackled by exact diagonalization with current architectures.
The correlation energies reported in Ref.~\onlinecite{Eriksen_2020} are gathered in Table \ref{tab:energy} alongside the best ph-AFQMC estimate from Ref.~\onlinecite{Lee_2020} based on a CAS(6,6) trial wave function.
The outcome of this work is nicely summarized in the abstract of Ref.~\onlinecite{Eriksen_2020}:
\textit{``In our assessment, the evaluated high-level methods are all found to qualitatively agree on a final correlation energy, with most methods yielding an estimate of the FCI value around $-863$ m$E_h$. However, we find the root-mean-square deviation of the energies from the studied methods to be considerable ($1.3$ m$E_h$), which in light of the acclaimed performance of each of the methods for smaller molecular systems clearly displays the challenges faced in extending reliable, near-exact correlation methods to larger systems.''}

\begin{table}
  \caption{
    The frozen-core correlation energy $\Delta E$ (in m$E_h$) of benzene in the cc-pVDZ basis set using various methods.
    \label{tab:energy}
  }
  \begin{ruledtabular}
    \begin{tabular}{llc}
      Method		&	\tabc{$\Delta E$}			&	Ref.	\\ 
      \hline
      ASCI			&	$-860.0$		&	\onlinecite{Eriksen_2020}	\\
      iCI			&	$-861.1$		&	\onlinecite{Eriksen_2020}	\\
      CCSDTQ		&	$-862.4$		&	\onlinecite{Eriksen_2020}	\\
      DMRG			&	$-862.8$		&	\onlinecite{Eriksen_2020}	\\
      FCCR			&	$-863.0$		&	\onlinecite{Eriksen_2020}	\\
      MBE-FCI		&	$-863.0$		&	\onlinecite{Eriksen_2020}	\\
      CAD-FCIQMC	&	$-863.4$		&	\onlinecite{Eriksen_2020}	\\
      AS-FCIQMC		&	$-863.7$		&	\onlinecite{Eriksen_2020}	\\
      SHCI			&	$-864.2$		&	\onlinecite{Eriksen_2020}	\\
      \hline
      ph-AFQMC		&	$-864.3(4)$		&	\onlinecite{Lee_2020}		\\
      \hline
      CIPSI			&	$-863.4$		&	This work					\\
   \end{tabular}
 \end{ruledtabular}
\end{table}

\begin{figure*}
	\includegraphics[width=0.4\linewidth]{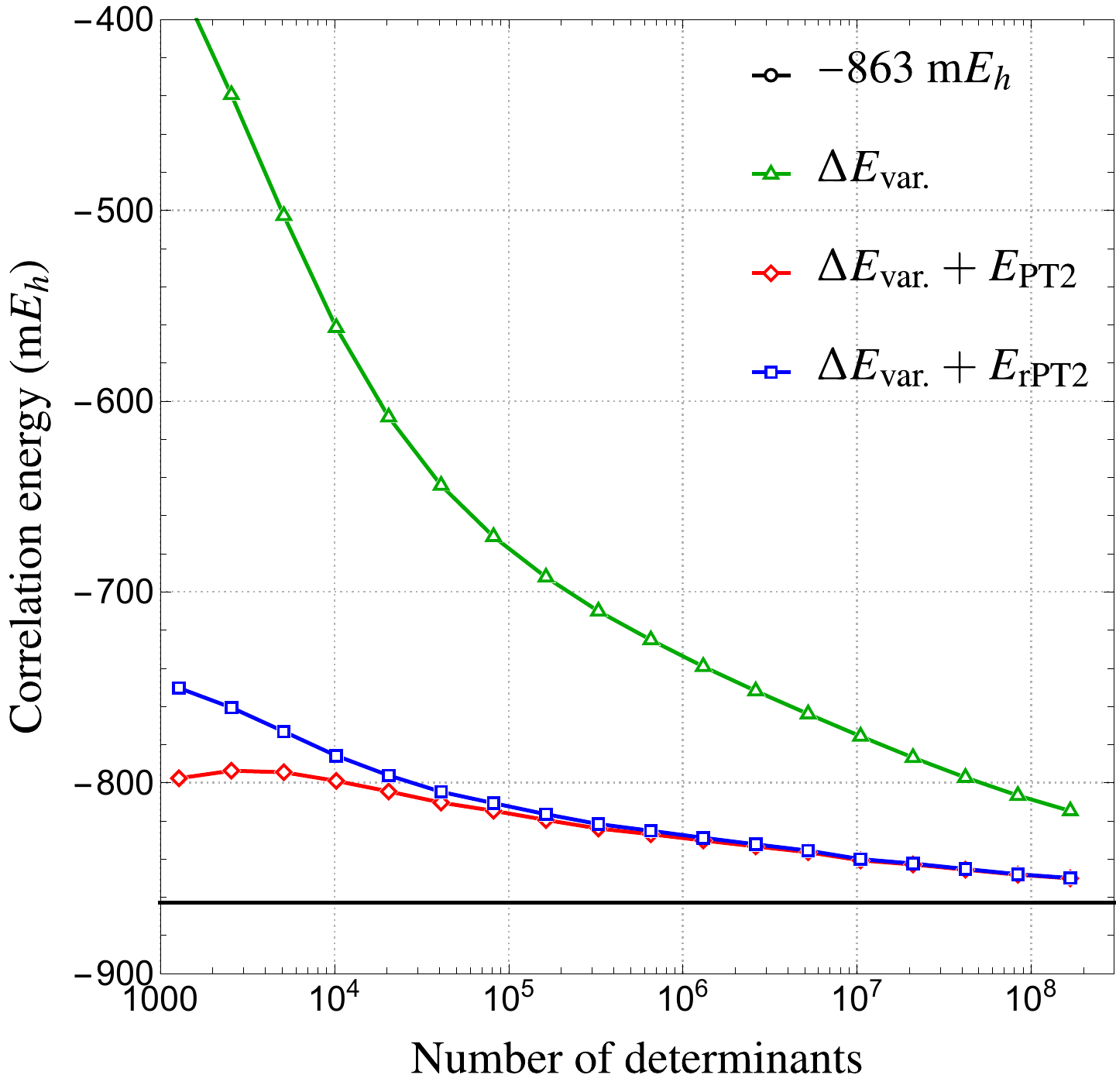}
	\hspace{0.08\linewidth}
	\includegraphics[width=0.4\linewidth]{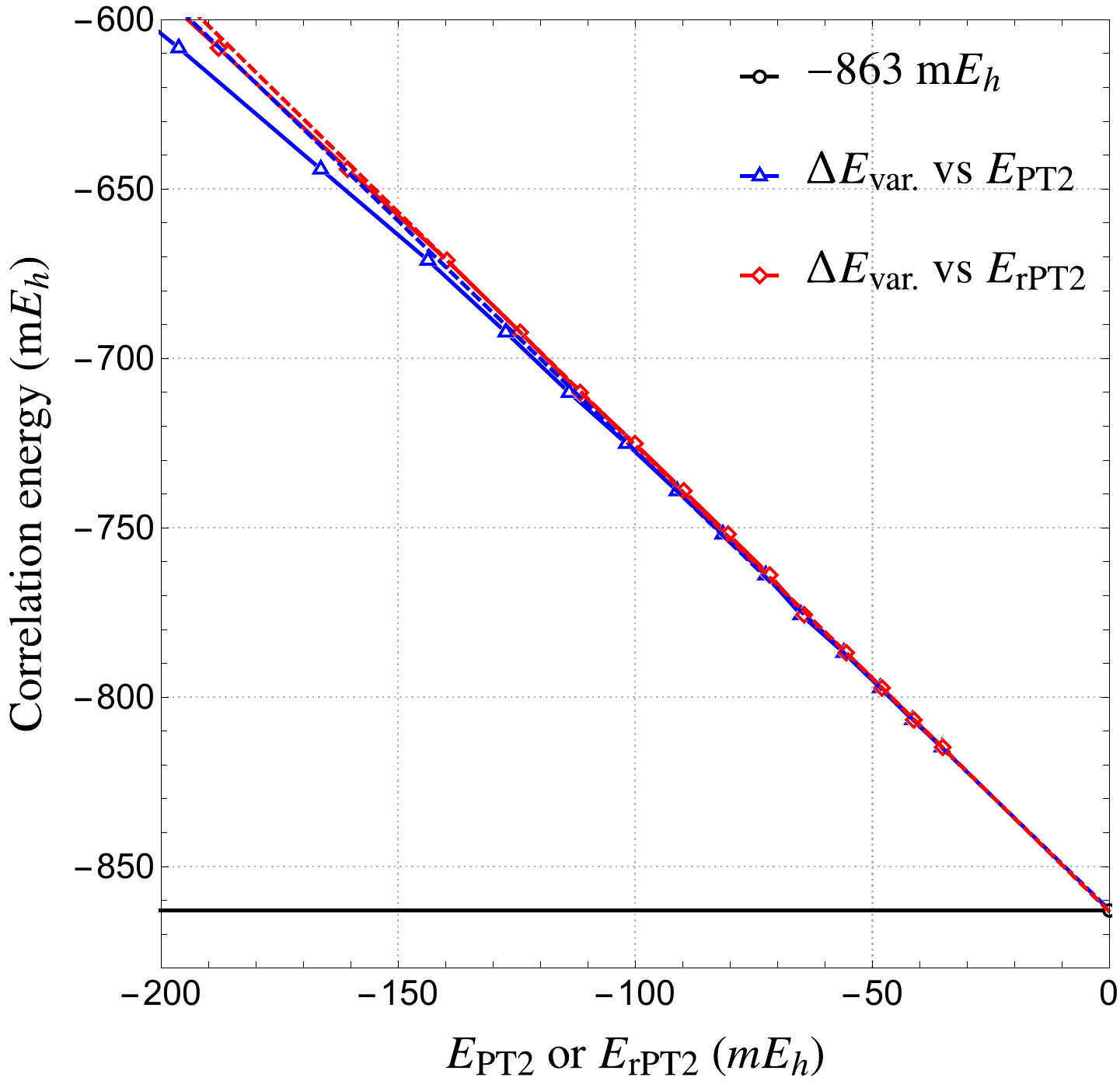}
	\caption{
	Convergence of the CIPSI correlation energy of benzene using localized orbitals.
	Left: $\Delta E_\text{var.}$, $\Delta E_\text{var.} + E_\text{PT2}$, and $\Delta E_\text{var.} + E_\text{rPT2}$ (in m$E_h$) as functions of the number of determinants in the variational space $N_\text{det}$.
	Right: $\Delta E_\text{var.}$ (in m$E_h$) as a function of $E_\text{PT2}$ or $E_\text{rPT2}$.
	The four-point linear extrapolation curves (dashed lines) are also reported.
	The theoretical estimate of $-863$ m$E_h$ from Ref.~\onlinecite{Eriksen_2020} is marked by a black line for comparison purposes.
	The statistical error bars associated with $E_\text{PT2}$ or $E_\text{rPT2}$ (not shown) are of the order of the size of the markers.
	\label{fig:CIPSI}
	}
\end{figure*}

\begin{table*}
\caption{Variational energy $E_\text{var.}$, second-order perturbative correction $E_\text{PT2}$ and its renormalized version $E_\text{rPT2}$ (in $E_h$) as a function of the number of determinants $N_\text{det}$ for the ground-state of the benzene molecule computed in the cc-pVDZ basis set. 
The statistical error on $E_\text{(r)PT2}$, corresponding to one standard deviation, are reported in parenthesis.}
\label{tab:NOvsLO}
\begin{ruledtabular}
\begin{tabular}{rcccccc}
		&	\mc{3}{c}{Natural orbitals}	&	\mc{3}{c}{Localized orbitals}	\\
		\cline{2-4} \cline{5-7}
\tabc{$N_\text{det}$}	&	\tabc{$E_\text{var.}$}	&	\tabc{$E_\text{var.}+E_\text{PT2}$}	&	\tabc{$E_\text{var.}+E_\text{rPT2}$}	
						&	\tabc{$E_\text{var.}$}	&	\tabc{$E_\text{var.}+E_\text{PT2}$}	&	\tabc{$E_\text{var.}+E_\text{rPT2}$}	\\	
\hline
          1\,280 & $-230.978\,056$  & $-231.559\,025(212)$  & $-231.463\,633(177)$    &   $-231.101\,676$  & $-231.519\,522(149)$   & $-231.472\,224(132)$  \\
          2\,560 & $-231.043\,712$  & $-231.542\,344(139)$  & $-231.474\,885(120)$    &   $-231.161\,264$  & $-231.515\,577(155)$   & $-231.482\,477(140)$  \\
          5\,120 & $-231.115\,142$  & $-231.534\,122(213)$  & $-231.488\,815(190)$    &   $-231.224\,632$  & $-231.516\,375(191)$   & $-231.495\,022(177)$  \\
         10\,240 & $-231.188\,813$  & $-231.531\,660(516)$  & $-231.502\,992(473)$    &   $-231.283\,295$  & $-231.520\,907(271)$   & $-231.507\,708(255)$  \\
         20\,480 & $-231.260\,065$  & $-231.534\,172(611)$  & $-231.517\,063(573)$    &   $-231.330\,209$  & $-231.526\,433(586)$   & $-231.518\,045(561)$  \\
         40\,960 & $-231.321\,906$  & $-231.538\,269(501)$  & $-231.528\,301(478)$    &   $-231.366\,008$  & $-231.532\,288(303)$   & $-231.526\,639(293)$  \\
         81\,920 & $-231.366\,895$  & $-231.541\,945(813)$  & $-231.535\,785(785)$    &   $-231.392\,888$  & $-231.536\,578(614)$   & $-231.532\,575(597)$  \\
        163\,840 & $-231.392\,866$  & $-231.545\,499(761)$  & $-231.541\,010(739)$    &   $-231.414\,132$  & $-231.541\,400(624)$   & $-231.538\,378(609)$  \\
        327\,680 & $-231.407\,802$  & $-231.548\,699(662)$  & $-231.544\,980(645)$    &   $-231.431\,952$  & $-231.545\,873(557)$   & $-231.543\,532(545)$  \\
        655\,360 & $-231.418\,752$  & $-231.551\,208(661)$  & $-231.548\,004(645)$    &   $-231.447\,007$  & $-231.548\,856(498)$   & $-231.547\,043(489)$  \\
     1\,310\,720 & $-231.428\,852$  & $-231.552\,760(616)$  & $-231.550\,006(603)$    &   $-231.460\,970$  & $-231.552\,137(453)$   & $-231.550\,723(446)$  \\
     2\,621\,440 & $-231.439\,324$  & $-231.553\,845(572)$  & $-231.551\,544(560)$    &   $-231.473\,751$  & $-231.555\,261(403)$   & $-231.554\,159(397)$  \\
     5\,242\,880 & $-231.450\,156$  & $-231.557\,541(534)$  & $-231.555\,558(524)$    &   $-231.485\,829$  & $-231.558\,303(362)$   & $-231.557\,451(358)$  \\
    10\,485\,760 & $-231.461\,927$  & $-231.559\,390(481)$  & $-231.557\,796(474)$    &   $-231.497\,515$  & $-231.562\,568(322)$   & $-231.561\,901(319)$  \\
    20\,971\,520 & $-231.474\,019$  & $-231.561\,315(430)$  & $-231.560\,063(424)$    &   $-231.508\,714$  & $-231.564\,707(275)$   & $-231.564\,223(273)$  \\
    41\,943\,040 & $-231.487\,978$  & $-231.564\,529(382)$  & $-231.563\,593(377)$    &   $-231.519\,122$  & $-231.567\,419(240)$   & $-231.567\,069(238)$  \\
    83\,886\,080 & $-231.501\,334$  & $-231.566\,994(317)$  & $-231.566\,325(314)$    &   $-231.528\,568$  & $-231.570\,084(199)$   & $-231.569\,832(198)$  \\
   167\,772\,160 & $-231.514\,009$  & $-231.569\,939(273)$  & $-231.569\,467(271)$    &   $-231.536\,655$  & $-231.571\,981(175)$   & $-231.571\,804(174)$  \\
  \end{tabular}  
\end{ruledtabular}  
\end{table*}  

\begin{table}
	\caption{Extrapolation distances, $\Delta E_{\text{dist}}$, defined as the difference between the final computed energy, $\Delta E_{\text{final}}$, and the extrapolated energy, $\Delta E_{\text{extrap.}}$ associated with ASCI, iCI, SHCI, DMRG, and CIPSI for the best blind-test and post-blind-test estimates of the correlation energy of benzene in the cc-pVDZ basis. 
	The final variational energies $\Delta E_{\text{var.}}$ are also reported.
	See Ref.~\onlinecite{Eriksen_2020} for more details.
	All correlation energies are given in m$E_h$.
	\label{tab:extrap_dist_table}
	}
	\begin{ruledtabular}
		\begin{tabular}{lcccc}
			Method & $\Delta E_{\text{var.}}$ & $\Delta E_{\text{final}}$ & $\Delta E_{\text{extrap.}}$ & $\Delta E_{\text{dist}}$ \\
			\hline
			\mc{4}{l}{Best blind-test estimates}								\\
			ASCI	&	$-737.1$	&	$-835.4$	&	$-860.0$	&	$-24.6$	\\
			iCI		&	$-730.0$	&	$-833.7$	&	$-861.1$	&	$-27.4$	\\
			SHCI	&	$-827.2$	&	$-852.8$	&	$-864.2$	&	$-11.4$	\\
			DMRG	&	$-859.2$	&	$-859.2$	&	$-862.8$	&	$-3.6$	\\
			\hline
			\mc{4}{l}{Best post-blind-test estimates}							\\
			ASCI	&	$-772.4$	&	$-835.2$	&	$-861.3$	&	$-26.1$	\\
			iCI		&	$-770.7$	&	$-842.8$	&	$-864.2$	&	$-21.3$	\\		
			SHCI	&	$-835.2$	&	$-854.9$	&	$-863.6$	&	$-8.7$	\\		
			\hline
			CIPSI	&	$-814.8$	&	$-850.2$	&	$-863.4$	&	$-13.2$	\\
		\end{tabular}
	\end{ruledtabular}
\end{table}

For the sake of completeness and our very own curiosity, we report in this Note the frozen-core correlation energy obtained with a fourth flavor of SCI known as \textit{Configuration Interaction using a Perturbative Selection made Iteratively} (CIPSI), \cite{Huron_1973} which also includes a PT2 correction.
In short, the CIPSI algorithm belongs to the family of SCI+PT2 methods.
The idea behind such methods is to slow down the exponential increase of the size of the CI expansion by retaining the most energetically relevant determinants only, thanks to the use of a second-order energetic criterion to select perturbatively determinants in the FCI space. 
However, performing SCI calculations rapidly becomes extremely tedious when one increases the system size as one hits the exponential wall inherently linked to these methods.

From a historical point of view, CIPSI is probably one of the oldest SCI algorithm. 
It was developed in 1973 by Huron, Rancurel, and Malrieu \cite{Huron_1973} (see also Ref.~\onlinecite{Evangelisti_1983}).
Recently, the determinant-driven CIPSI algorithm has been efficiently implemented \cite{Giner_2013,Giner_2015} in the open-source programming environment {\QP} by our group enabling to perform massively parallel computations. \cite{Garniron_2017,Garniron_2018,Garniron_2019}
In particular, we were able to compute highly-accurate ground- and excited-state energies for small- and medium-sized molecules (including benzene). \cite{Loos_2018a,Loos_2019,Loos_2020a,Loos_2020b,Loos_2020c}
CIPSI is also frequently used to provide accurate trial wave function for QMC calculations. \cite{Caffarel_2014,Caffarel_2016a,Caffarel_2016b,Giner_2013,Giner_2015,Scemama_2015,Scemama_2016,Scemama_2018,Scemama_2018b,Scemama_2019,Dash_2018,Dash_2019} 
The particularity of the current implementation is that the selection step and the PT2 correction are computed \textit{simultaneously} via a hybrid semistochastic algorithm \cite{Garniron_2017,Garniron_2019} (which explains the statistical error associated with the PT2 correction in the following).
Moreover, a renormalized version of the PT2 correction (dubbed rPT2 below) has been recently implemented and tested for a more efficient extrapolation to the FCI limit thanks to a partial resummation of the higher-order of perturbation. \cite{Garniron_2019} 
We refer the interested reader to Ref.~\onlinecite{Garniron_2019} where one can find all the details regarding the implementation of the rPT2 correction and the CIPSI algorithm.

Being late to the party, we obviously cannot report blindly our CIPSI results.
However, following the philosophy of Eriksen \textit{et al.} \cite{Eriksen_2020} and Lee \textit{et al.}, \cite{Lee_2020} we will report our results with the most neutral tone, leaving the freedom to the reader to make up his/her mind.
We then follow our usual ``protocol'' \cite{Scemama_2018,Scemama_2018b,Scemama_2019,Loos_2018a,Loos_2019,Loos_2020a,Loos_2020b,Loos_2020c} by performing a preliminary SCI calculation using Hartree-Fock orbitals in order to generate a SCI wave function with at least $10^7$ determinants.
Natural orbitals are then computed based on this wave function, and a new SCI calculation is performed with this new natural set of orbitals. 
This has the advantage to produce a smoother and faster convergence of the SCI energy toward the FCI limit.
The total SCI energy is defined as the sum of the variational energy $E_\text{var.}$ (computed via diagonalization of the CI matrix in the reference space) and a second-order perturbative correction $E_\text{(r)PT2}$ which takes into account the external determinants, \ie, the determinants which do not belong to the variational space but are linked to the reference space via a nonzero matrix element. The magnitude of $E_\text{(r)PT2}$ provides a qualitative idea of the ``distance'' to the FCI limit.
As mentioned above, SCI+PT2 methods rely heavily on extrapolation, especially when one deals with medium-sized systems.
We then linearly extrapolate the total SCI energy to $E_\text{(r)PT2} = 0$ (which effectively corresponds to the FCI limit). 
Note that, unlike excited-state calculations where it is important to enforce that the wave functions are eigenfunctions of the $\Hat{S}^2$ spin operator, \cite{Applencourt_2018} the present wave functions do not fulfil this property as we aim for the lowest possible energy of a singlet state. 
We have found that $\expval*{\Hat{S}^2}$ is, nonetheless, very close to zero ($\sim 5 \times 10^{-3}$ a.u.).
The corresponding energies are reported in Table \ref{tab:NOvsLO} as functions of the number of determinants in the variational space $N_\text{det}$.

A second run has been performed with localized orbitals. 
Starting from the same natural orbitals, a Boys-Foster localization procedure \cite{Boys_1960} was performed in several orbital windows: i) core, ii) valence $\sigma$, iii) valence $\pi$, iv) valence $\pi^*$, v) valence $\sigma^*$, vi) the higher-lying $\sigma$ orbitals, and vii) the higher-lying $\pi$ orbitals. 
\footnote{Indices of molecular orbitals for Boys-Foster localization procedure:
core [1--6];
$\sigma$ [7--18];
$\pi$ [19--21]; 
$\pi^*$ [22--24];
$\sigma^*$ [25--36];
higher-lying $\pi$ [39,41--43,46,49,50,53--57,71--74,82--85,87,92,93,98];
higher-lying $\sigma$ [37,38,40,44,45,47,48,51,52,58--70,75--81,86,88--91,94--97,99--114].}
Like Pipek-Mezey, \cite{Pipek_1989} this choice of orbital windows allows to preserve a strict $\sigma$-$\pi$ separation in planar systems like benzene.
As one can see from the energies of Table \ref{tab:NOvsLO}, for a given value of $N_\text{det}$, the variational energy as well as the PT2-corrected energies are much lower with localized orbitals than with natural orbitals. 
Indeed, localized orbitals significantly speed up the convergence of SCI calculations by taking benefit of the local character of electron correlation.\cite{Angeli_2003,Angeli_2009,BenAmor_2011,Suaud_2017,Chien_2018,Eriksen_2020}
We, therefore, consider these energies more trustworthy, and we will base our best estimate of the correlation energy of benzene on these calculations. 
The convergence of the CIPSI correlation energy using localized orbitals is illustrated in Fig.~\ref{fig:CIPSI}, where one can see the behavior of the correlation energy, $\Delta E_\text{var.}$ and $\Delta E_\text{var.} + E_\text{(r)PT2}$, as a function of $N_\text{det}$ (left panel).
The right panel of Fig.~\ref{fig:CIPSI} is more instructive as it shows $\Delta E_\text{var.}$ as a function of $E_\text{(r)PT2}$, and their corresponding four-point linear extrapolation curves that we have used to get our final estimate of the correlation energy. 
(In other words, the four largest variational wave functions are considered to perform the linear extrapolation.)
From this figure, one clearly sees that the rPT2-based correction behaves more linearly than its corresponding PT2 version, and is thus systematically employed in the following.

Our final number are gathered in Table \ref{tab:extrap_dist_table}, where, following the notations of Ref.~\onlinecite{Eriksen_2020}, we report, in addition to the final variational energies $\Delta E_{\text{var.}}$, the
extrapolation distances, $\Delta E_{\text{dist}}$, defined as the difference between the final computed energy, $\Delta E_{\text{final}}$, and the extrapolated energy, $\Delta E_{\text{extrap.}}$ associated with ASCI, iCI, SHCI, DMRG, and CIPSI. 
The three flavours of SCI fall into an interval ranging from $-860.0$ m$E_h$ (ASCI) to $-864.2$ m$E_h$ (SHCI), while the other non-SCI methods yield correlation energies ranging from $-863.7$ to $-862.8$ m$E_h$ (see Table \ref{tab:energy}). Our final CIPSI number (obtained with localized orbitals and rPT2 correction via a four-point linear extrapolation) is $-863.4(5)$ m$E_h$, where the error reported in parenthesis represents the fitting error (not the extrapolation error for which it is much harder to provide a theoretically sound estimate).
\footnote{Using the last 3, 4, 5, and 6 largest wave functions to perform the linear extrapolation yield the following correlation energy estimates: $-863.1(11)$, $-863.4(5)$, $-862.1(8)$, and $-863.5(11)$ mE$_h$, respectively. 
These numbers vary by $1.4$ mE$_h$. 
The four-point extrapolated value of $-863.4(5)$ mE$_h$ that we have chosen to report as our best estimate corresponds to the smallest fitting error. 
Quadratic fits yield much larger variations and are discarded in practice.
Due to the stochastic nature of $E_\text{rPT2}$, the fifth point is slightly off as compared to the others. 
Taking into account this fifth point yield a slightly smaller estimate of the correlation energy [$-862.1(8)$ mE$_h$], while adding a sixth point settles down the correlation energy estimate at $-863.5(11)$ mE$_h$
}
For comparison, the best post blind test SHCI estimate is $-863.3$ m$E_h$, which agrees almost perfectly with our best CIPSI estimate, while the best post blind test ASCI and iCI correlation energies are $-861.3$ and $-864.15$ m$E_h$, respectively (see Table \ref{tab:extrap_dist_table}).

The present calculations have been performed on the AMD partition of GENCI's Irene supercomputer. 
Each Irene's AMD node is a dual-socket AMD Rome (Epyc) CPU@2.60 GHz with 256GiB of RAM, with a total of 64 physical CPU cores per socket.
These nodes are connected via Infiniband HDR100.
The first step of the calculation, \ie, performing a CIPSI calculation up to $N_\text{det} \sim 10^7$ with Hartree-Fock orbitals in order to produce natural orbitals, takes roughly 24 hours on a single node, and reaching the same number of determinants with natural orbitals or localized orbitals takes roughly the same amount of time. 
A second 24-hour run on 10 distributed nodes was performed to push the selection to $8 \times 10^7$ determinants, and a third distributed run using 40 nodes was used to reach $16 \times 10^7$ determinants.
In total, the present calculation has required 150k core hours, most of it being spent in the last stage of the computation.

We thank Janus Eriksen and Cyrus Umrigar for useful comments.
This work was performed using HPC resources from GENCI-TGCC (2020-gen1738) and from CALMIP (Toulouse) under allocation 2020-18005.
PFL and AS have received funding from the European Research Council (ERC) under the European Union's Horizon 2020 research and innovation programme (Grant agreement No.~863481).

The data that support the findings of this study are openly available in Zenodo at \href{http://doi.org/10.5281/zenodo.4075286}{http://doi.org/10.5281/zenodo.4075286}.

\end{document}